\documentclass[aps,prb,reprint,superscriptaddress,longbibliography,floatfix]{revtex4-2}

\usepackage{graphicx}
\usepackage{amsmath}
\usepackage{amssymb}
\usepackage{booktabs} 
\usepackage{tikz}
\usetikzlibrary{arrows.meta, positioning, decorations.pathmorphing}
\usepackage{float}
\usepackage[dvipsnames]{xcolor} \usepackage[colorlinks=true,linkcolor=blue,citecolor=blue,urlcolor=blue]{hyperref}

\usepackage[normalem]{ulem}
\definecolor{Gray}{rgb}{0,0,0}

\begin{document}

\title{Controlling Waiting Time Statistics in Monitored Collective Spins: \\ Mitigating Detector's Resolution Barrier in Measurement-Induced Phase Transitions}

\author{Tanbir Islam}
\email{tanbir\_i@id.uff.br}
\affiliation{Instituto de F\'{i}sica, Universidade Federal Fluminense, Av. Gal. Milton Tavares de Souza s/n, Gragoat\'{a}, 24210-346 Niter\'{o}i, Rio de Janeiro, Brazil}

\author{Fernando Iemini}
\affiliation{Instituto de F\'{i}sica, Universidade Federal Fluminense, Av. Gal. Milton Tavares de Souza s/n, Gragoat\'{a}, 24210-346 Niter\'{o}i, Rio de Janeiro, Brazil}

\date{\today}

\begin{abstract}
In collective dissipative spin systems, the postselection barrier can be partially mitigated; however, a further obstacle may be posed by the finite temporal resolution of detectors. In this work, we investigate how initial-state inhomogeneities can control waiting-time statistics between quantum jumps, thereby mitigating the detector-resolution problem. We consider a collectively monitored spin model with a boundary time-crystalline phase, introducing inhomogeneity by partitioning the ensemble into two subsystems rotated by an angle $\theta$. We find that the measurement-induced phase transition survives under inhomogeneities, with distinct entanglement scaling regimes. The waiting time increases with $\theta$, scaling as $1/N$ but with a prefactor strongly enhanced by orders of magnitude, and in the anti-aligned limit $\theta = \pi$ it remains finite, fully resolving the resolution barrier. This mitigation, however, comes at a cost: the entanglement saturation time becomes significantly longer, partially reintroducing the postselection barrier. Our results highlight a trade-off between detector resolution and postselection overhead, with direct implications for the experimental observation of measurement-induced phenomena.
\end{abstract}

\maketitle

\section{Introduction}
\label{sec:introduction}

The dynamics of a quantum many-body system under continuous observation can host a rich phenomenology, as external measurements can alter the dynamics in ways that are invisible at the level of the ensemble-averaged density matrix \cite{Fazio2025_ManyBodyOpen, Passarelli2024_ManybodyDynamics}. In quantum many-body systems, the interplay between coherent evolution and measurements can qualitatively reshape quantum correlations, inducing a transition that remains completely hidden at the level of the average density matrix \cite{Li2018_QuantumZeno, Skinner2019_Measurementinduced}. Resolving the system’s evolution conditioned on specific measurement outcomes provides access to individual quantum trajectories. This access is essential for computing observables that are non-linear in the conditional density matrix $\hat{\rho}_c$~\cite{Carmichael1999, Wiseman2009, Jacobs2014}, most notably the von Neumann entanglement entropy, $S=-\mathrm{Tr}(\hat{\rho}_c\ln\hat{\rho}_c)$, and the purity, $P=\mathrm{Tr}(\hat{\rho}_c^2)$, which serve as trajectory-level diagnostics of measurement-induced phase transitions (MIPTs)~\cite{Fisher2023Review}. A large body of subsequent work has further explored measurement-induced phases in monitored quantum systems through the analysis of individual quantum trajectories across different settings \cite{Li2019_PRB, Szyniszewski2019_PRB, Jian2020_PRB, Zabalo2020_PRB, Szyniszewski2020_PRL, Turkeshi2020_PRB, Lunt2021_PRB, Sierant2022_PRB, Nahum2021_PRXQ, Zabalo2022_PRL, Sierant2022_PRL, Chiriaco2023_PRB, Klocke2023_PRX, Cao2019_SciPostPhys, Nahum2020_PRRes, Buchhold2021_PRX, Jian2022_PRB, Coppola2022_PRB, Fava2023_PRX, Poboiko2023_PRX, Merritt2023_PRB, Alberton2021_PRL, Turkeshi2021_PRB, Turkeshi2022_PRB_L, Piccitto2022_PRB, Piccitto2023_SciPostCore, Tirrito2023_SciPost, Paviglianiti2023_PRB, Rossini2020_PRB, Tang2020_PRRes, Fuji2020_PRB, Sierant2022_Quantum, Doggen2022_PRRes, Altland2022_PRRes, Jian2023_PRB}.

Despite intense theoretical activity, experimental studies of measurement-induced phase transitions~\cite{Noel2022, Koh2023} are very limited and remain severely constrained by the 
post-selection problem~\cite{leung2023theoryfreefermionsdynamics, Garratt_2024, Passarelli2024_ManybodyDynamics}. The post-selection problem is a fundamental obstacle to observing monitored phases. Characterizing observables or entanglement along a single quantum 
trajectory requires reproducing the same sequence of measurement outcomes multiple times. However, since the number of possible trajectories grows exponentially with the number of recorded jump 
events, the probability of repeatedly realizing a given trajectory becomes exponentially small~\cite{Passarelli2024_ManybodyDynamics}. More precisely, a trajectory can be represented as a binary string of jump events occurring up to the saturation time $\tau_{\rm sat}$, 
so that $P_{\rm same} \propto 2^{-\tau_{\rm sat}}$. In generic monitored many-body systems with local measurements, $\tau_{\rm sat}$ grows extensively with system size, $\tau_{\rm sat} \sim \mathcal{O}(N)$, making the post-selection overhead exponential in $N$ and severely 
limiting experimental access to large systems. This barrier has limited experimental explorations of MIPTs to small systems and represents a fundamental obstacle to scaling up these studies~\cite{GoogleQuantumAI2023_Measurementinduced}.

\begin{figure*}
    \centering
    \includegraphics[scale=0.225]{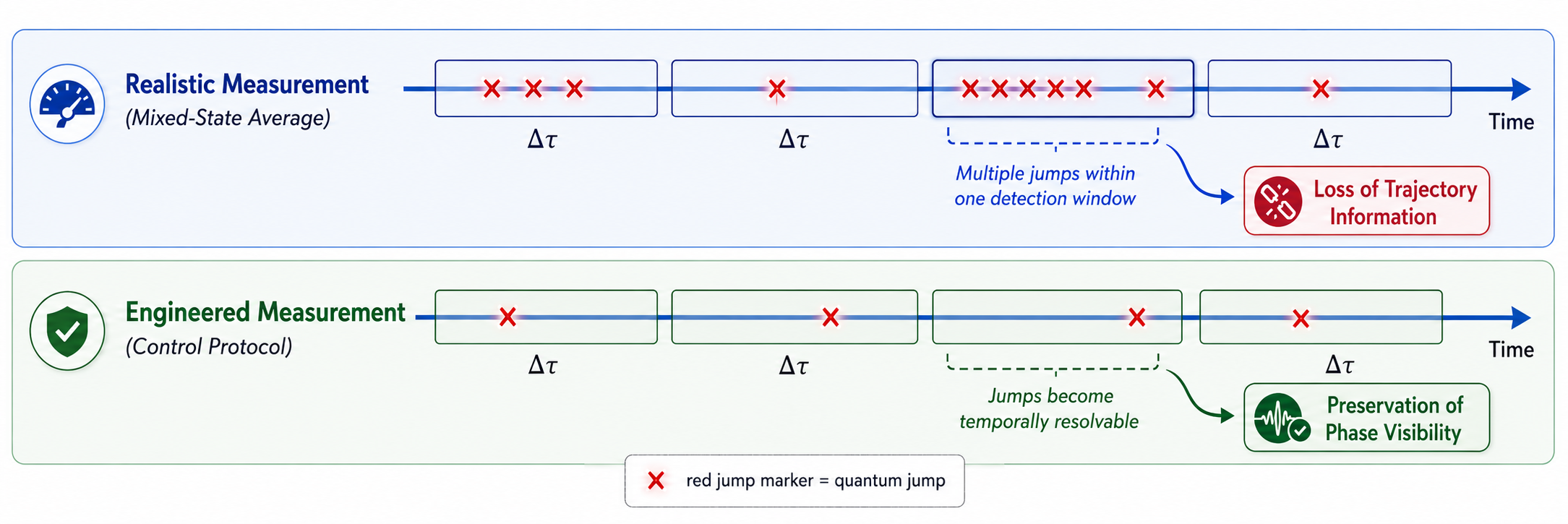}
    \caption{\textbf{ Schematic of the detector-resolution problem and control protocol.} \textbf{(Top panel)} For a finite detector resolution $\Delta\tau$, multiple quantum jumps may occur within the same time bin, making distinct trajectories experimentally indistinguishable and leading to a mixed-state description. \textbf{(Bottom panel)} The engineered initial-state protocol suppresses the jump rate, increasing the waiting time between events so that individual jumps remain resolvable within each bin. }
    \label{fig:jump}
\end{figure*}

Recent efforts to address the post-selection barrier have followed different routes. 
Some approaches aim to bypass direct trajectory post-selection by extracting trajectory-dependent information through classical post-processing, modified measurement schemes, feedback protocols, spacetime-duality constructions, or related circuit-based strategies~\cite{Li2023,Garratt_2024,Gullans2020_Scalableprobes,Ippoliti2021_PostselectionFree,Iadecola2023,Buchhold2022_arXiv}. 
These methods can make nonlinear trajectory observables accessible in specific settings, but they do not  reduce the post-selection cost in the same sense. 

A more direct form of mitigation was identified in monitored infinite-range systems, where the entanglement dynamics saturates on a timescale that grows only logarithmically in $N$~\cite{Passarelli2024_ManybodyDynamics,Delmonte2024_Measurementinduced,Li2024_MonitoredLongRange,Li2025_EmergentDeterministic}. 
In this case, the post-selection barrier is not removed, but the scaling of the required overhead can be reduced from exponential to polynomial in favorable regimes. 
However, practical implementation still faces an intrinsic detector-resolution problem, where the finite time resolution of experimental devices forces a coarse-graining of quantum trajectories, degrading pure-state evolution into a mixed-state description~\cite{Delmonte2024_Measurementinduced}. 
This degradation is governed by the competition between the detector's finite temporal resolution, $\Delta \tau$, and the characteristic waiting time between quantum jumps, $W$. 
Usually, in collective systems $W$ decreases with system size, pushing the dynamics into a regime where $\Delta \tau \gtrsim W$. 
Consequently, multiple emission events occurring within a single detection bin become experimentally indistinguishable, as illustrated in Fig.~\ref{fig:jump}. 
This unavoidable coarse-graining requires averaging over microscopically distinct trajectories, which degrades the purity of the conditional state and obscures the fine-grained entanglement correlations needed to identify measurement-induced phase transitions. 
The detector resolution therefore acts as a critical bottleneck for observing trajectory-dependent physics in large-scale quantum simulators. 

In this work, we propose a mitigation of the detector-resolution limit by engineering the initial state of the system. We consider a collectively monitored spin model that exhibits a boundary time-crystalline phase and a postselection-free character in the homogeneous limit. By partitioning the ensemble into two subsystems rotated by an angle $\theta$, we introduce a controllable degree of inhomogeneity. We demonstrate that appropriate initial-state preparation can suppress the collective decay rate, increasing the waiting time between quantum jumps and thereby partially resolving the detector-resolution barrier. Our analysis reveals a nuanced picture though. While the waiting time increases significantly with the inhomogeneity angle, and in the extreme anti-aligned case it remains finite and independent of system size, thus fully resolving the detector-resolution problem, it comes at a cost: the entanglement saturation time, which in the homogeneous case grows only logarithmically with system size, becomes significantly longer in the presence of inhomogeneities, thereby partially reintroducing the postselection barrier.
Our results thus highlight a fundamental trade-off between detector resolution and postselection overhead, and provide a pathway for controlling waiting time statistics through initial-state engineering, with direct implications for the experimental observation of measurement-induced phenomena.

This manuscript is organized as follows. In Sec.~\ref{sec:model} we introduce the monitored collective spin model, discuss its phase diagram, and describe the quantum-trajectory framework. In Sec.~\ref{inhom} we present the inhomogeneous initial-state preparation protocol. In Sec.~\ref{sec:results} we present our results for the waiting time statistics and entanglement dynamics. Finally, in Sec.~\ref{sec:Conclusion} we conclude and discuss perspectives.

\section{The Model}
\label{sec:model}

\subsection{Collective Spin Model}

We consider a collectively driven and monitored ensemble of $N$ two-level atoms. 
The ensemble-averaged density matrix evolves according to the Lindblad master equation
\begin{equation}
\dot{\hat{\rho}}
=
\mathcal{L}(\hat{\rho})
=
-i[\hat{H},\hat{\rho}]
+
\hat{L}\hat{\rho}\hat{L}^{\dagger}
-
\frac{1}{2}
\left\{
\hat{L}^{\dagger}\hat{L},\hat{\rho}
\right\}.
\label{eq:lindblad_general}
\end{equation}
The coherent part of the dynamics is generated by
\begin{equation}
\hat{H}=\omega_0 \hat{J}_x ,
\label{eq:Hamiltonian}
\end{equation}
where $\omega_0$ is the drive strength,
with corresponding collective spin operators
$\hat{J}_\alpha=\frac{1}{2}\sum_{i=1}^{N}\hat{\sigma}_i^\alpha $, 
where  $\hat{\sigma}_i^\alpha$ $(\alpha=x,y,z)$ are  Pauli operators acting on the $i$-th atom.
The system is also coupled to a monitored environment through a collective decay channel. This coupling is described by the jump operator
\begin{equation}
\hat{L}=\sqrt{\frac{2\kappa}{N}}\,\hat{J}_- ,
\label{eq:jump_operator}
\end{equation}
where the collective spin operators are defined as $\hat{J}_\pm=\hat{J}_x \pm i \hat{J}_y$, which are the collective raising and lowering operators and $\kappa$ is the collective decay rate. The factor $1/\sqrt{N}$ ensures a well-defined thermodynamic scaling of the dissipative dynamics. Suppression of collective decay through such interference effects has been directly observed experimentally in clouds of cold atoms~\cite{Guerin2016_Subradiance}, and is closely related to the segregation of fast and slow decay channels found more generally in open many-body systems described by an effective non-Hermitian Hamiltonian~\cite{Volya_2003}. A key structural property of the collective model is that the Lindbladian dynamics does not mix different total-spin irreducible representations. Indeed, one has that both $[\hat J^2,H]=[\hat J^2,\hat L]=0$ and the total angular momentum $\hat J^2=\hat{J}_x^2+\hat{J}_y^2+\hat{J}_z^2$,  with quantum numbers $J^2 = J(J+1)$ where $J=0,...,N/2$, is conserved along the ensemble-averaged dynamics. This strong symmetry is what our control protocol exploits: it is specific to the all-to-all structure of the model and has no direct analogue in systems with local interactions, where no operator plays the role of $\hat J^2$.

The model governed by Eq.~\eqref{eq:lindblad_general} supports two distinct dynamical regimes separated by a collective dissipative transition controlled by the drive-to-dissipation ratio $\omega_0/\kappa$. 
For homogeneous initial states, where the dynamics is confined to the fully symmetric Dicke sector with total spin $J=N/2$, the critical point occurs at $(\omega_0/\kappa)_c=1$~\cite{Iemini2018_BoundaryTimeCrystals}. 
Below this threshold, $\omega_0/\kappa<1$, the system relaxes to a trivial time-independent steady state characterized by static collective magnetization. 
Above it, $\omega_0/\kappa>1$, the ensemble-averaged dynamics displays persistent oscillations of collective observables such as the magnetization, corresponding to a boundary time-crystalline regime. 

\begin{figure}
\includegraphics[scale=0.43]{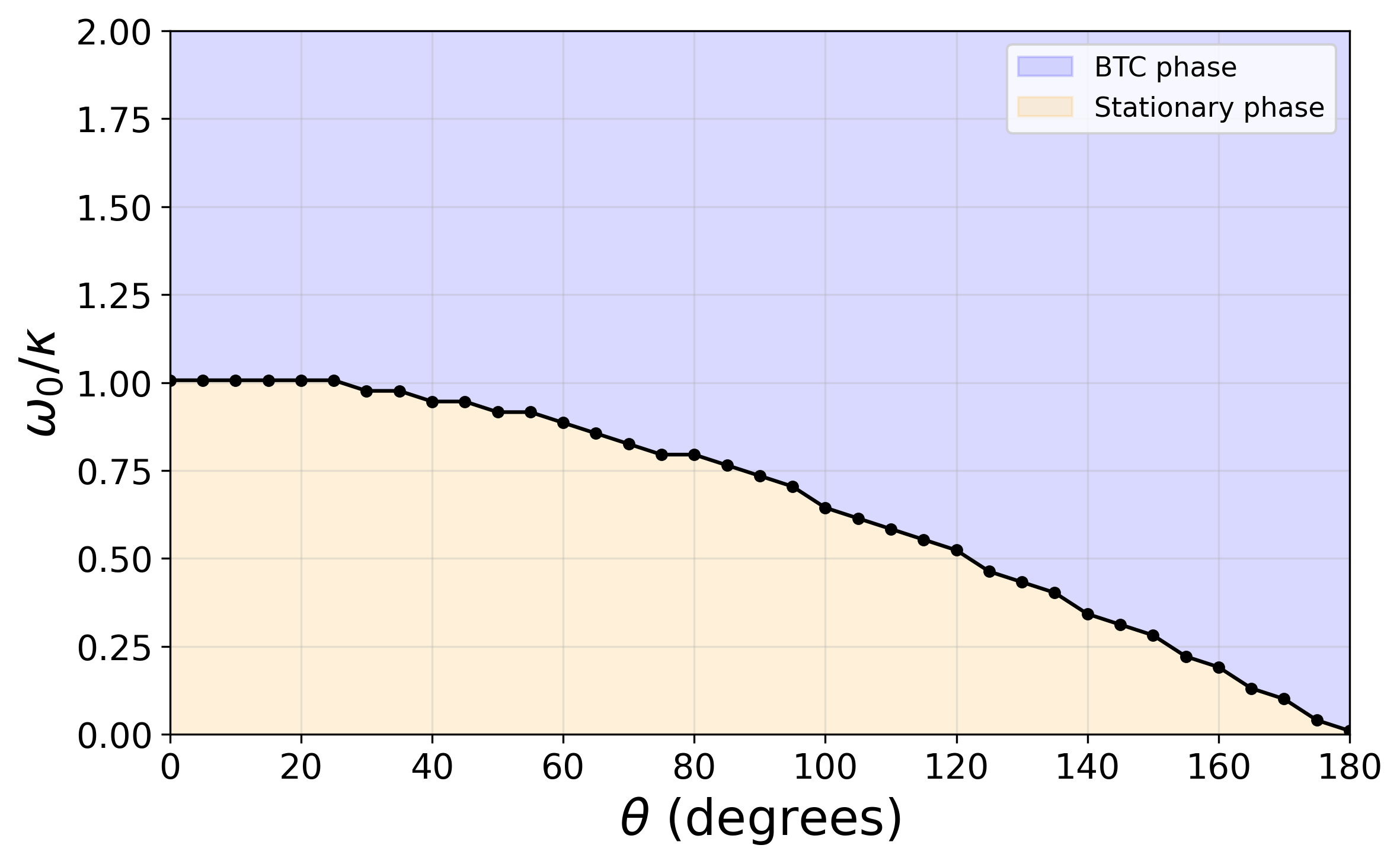}
\caption{Mean-field phase diagram of the collective spin model as a function of the initial-state inhomogeneity angle $\theta$. The black curve shows the critical drive-to-dissipation ratio $(\omega_0/\kappa)_c$ below which the ensemble-averaged dynamics settles into a stationary phase (orange region), and above which it supports persistent oscillations in a boundary time-crystalline (BTC) phase (blue region). 
}
\label{fig:mf_phase_diagram}
\end{figure}

\begin{figure*}
\begin{minipage}[t]{0.48\textwidth}
  \raggedright (a)\\[2pt]
  \includegraphics[width=\textwidth]{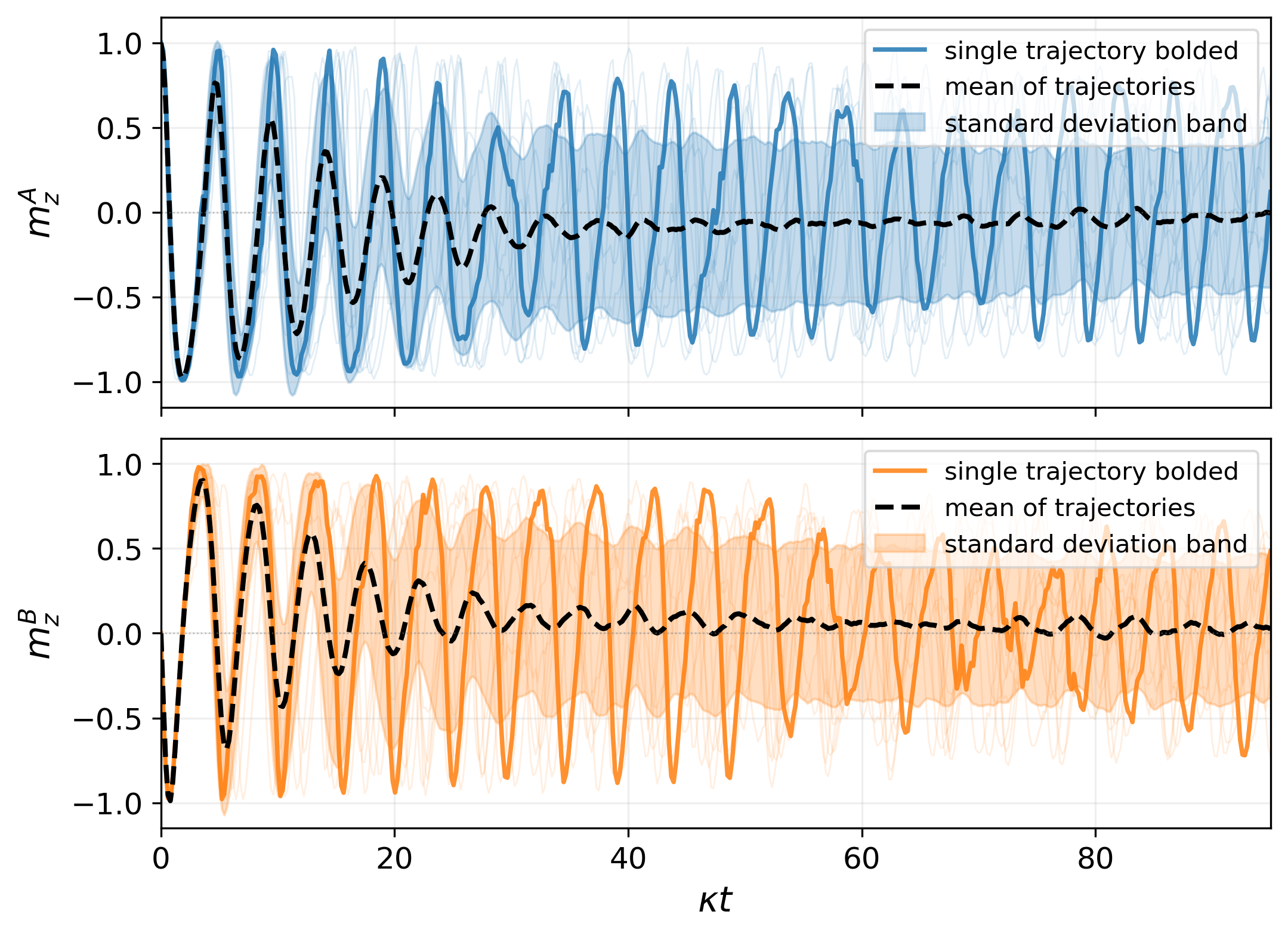}
\end{minipage}
\hfill
\begin{minipage}[t]{0.48\textwidth}
  \raggedright (b)\\[2pt]
  \includegraphics[width=\textwidth]{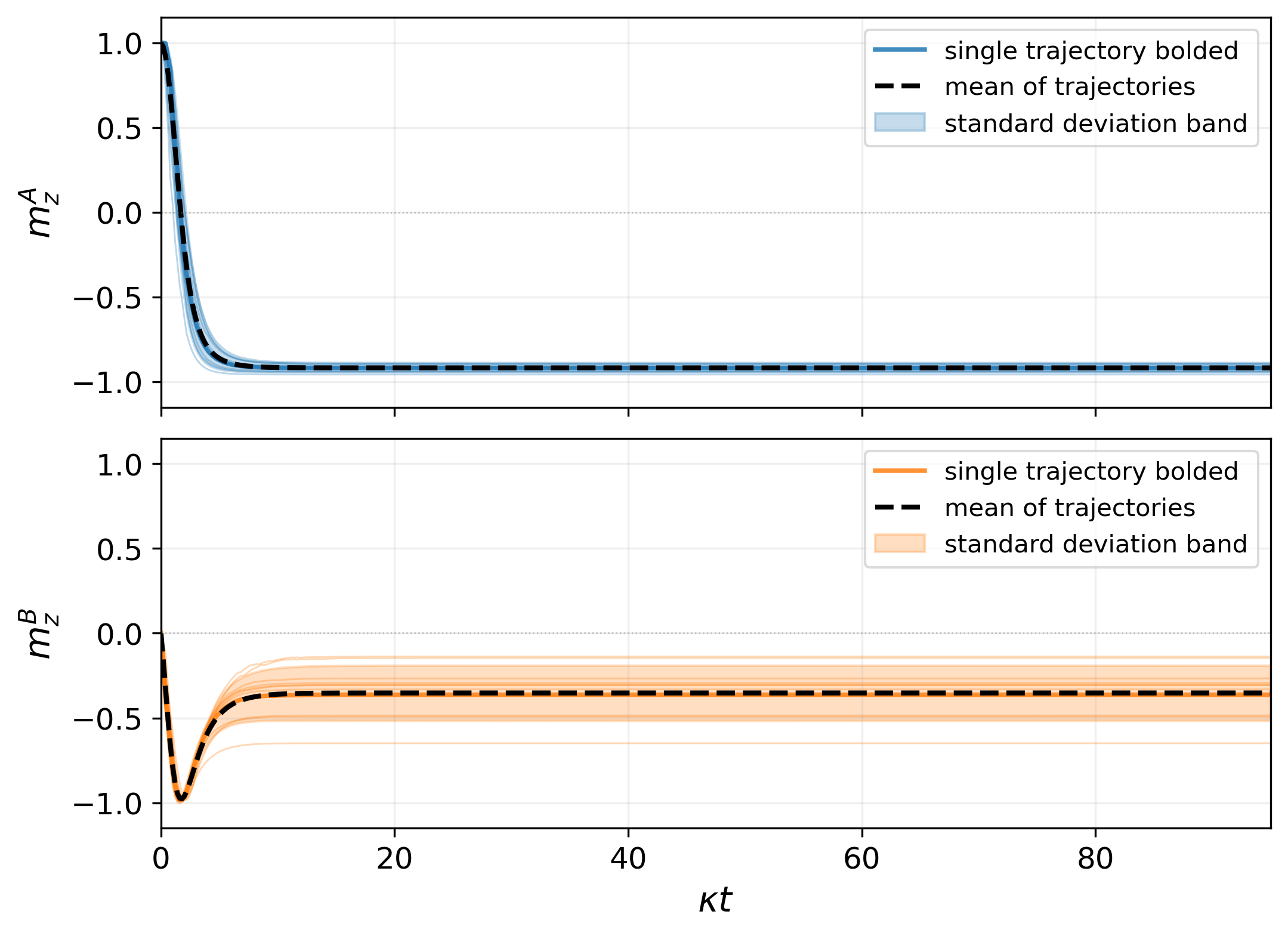}
\end{minipage}
\caption{{\textbf{(a)} Sub-ensemble magnetizations $m_z^{A}$ (blue, top sub-panel) and $m_z^{B}$ (orange, bottom sub-panel) along quantum trajectories for $N=60$ and $\theta=90^\circ$ in the BTC phase ($\omega_0/\kappa=1.5$). Dashed black lines show the trajectory-averaged mean, which decays toward zero due to random phase cancellation across trajectories; shaded regions indicate one standard deviation across $250$ trajectories; bold colored lines show a single representative trajectory; faint lines show ten further individual trajectories. The magnetizations oscillate persistently over the whole time window, with the random phase of each realization reflected in the large standard deviation. \textbf{(b)} Same quantities in the stationary phase ($\omega_0/\kappa=0.3$), where both sub-ensembles relax to fixed values by $\kappa t\approx10$, and the standard deviation across trajectories becomes small and time independent.}
}
\label{fig:mf_phase_diagram_1}
\end{figure*}

In the case of  inhomogeneous initial states, which break global permutation symmetry and populate lower total-spin sectors, the phase transition of the model  is shifted to smaller values of the drive-to-dissipation ratio, $(\omega_0/\kappa)_c<1$~\cite{Iemini_2024_Dynamics}.
 Considering a bi-partition of the full ensemble, rotated by an angle of $\theta$ from each other (as discussed in detail in Sec.\ref{inhom}), we compute the phase diagram within a mean-field approximation for the average magnetisation - as shown in Fig.~\ref{fig:mf_phase_diagram}. This mean-field treatment is justified by a rigorous proof of the exactness of semiclassical equations for open Dicke-type models with all-to-all couplings in the thermodynamic limit~\cite{Carollo_2021}, and driven-dissipative phase transitions of this general type have recently been realized experimentally in driven atomic ensembles~\cite{Ferioli_2023}. The phase boundary is determined numerically by integrating the mean-field equations of motion till the long time regime and the corresponding standard deviations for each sub-ensemble magnetization. In the stationary phase both sub-ensembles magnetization relax to fixed points with vanishing standard deviations, while in the BTC phase there are persistent oscillations with large standard deviations - as illustrated at the trajectory level in Fig.~\ref{fig:mf_phase_diagram_1}. The critical point $(\omega_0/\kappa)_c$ is the smallest $\omega_0/\kappa$ at which the standard deviation exceeds a threshold of $0.05$.  As $\theta$ increases from $0$ to $\pi$, the boundary of the BTC phase moves to progressively smaller drive-to-dissipation ratios, meaning that a weaker drive suffices to sustain persistent oscillations once the initial state is sufficiently inhomogeneous. Within the parameter range explored here the order parameter vanishes smoothly at $(\omega_0/\kappa)_c(\theta) = \cos(\theta/2)$ (derived in Appendix~\ref{app:meanfield}), consistent with a continuous transition for  all $\theta < \pi$.

\subsection{Quantum Trajectories}

The quantum-jump (QJ) unraveling of the Lindblad master equation generates an ensemble of stochastic pure-state trajectories, uniquely specified by the random sequence of jump events and their occurrence times~\cite{Dalibard1992_WaveFunction,Carmichael1999,Wiseman2009,Plenio1998_QuantumJump,Daley2014_QuantumTrajectories}. While the corresponding Lindblad master equation~\eqref{eq:lindblad_general} accurately reproduces the ensemble-averaged dynamics~\cite{Wiseman2009}, it is intrinsically insensitive to higher-order statistical properties of the trajectory distribution, i.e. non-linear observables along the quantum state trajectories.

Within the quantum-jump unraveling, the conditional state evolves under an effective non-Hermitian Hamiltonian,
\begin{equation}
\hat{H}_{\mathrm{eff}}
=
\hat{H}
-
i\frac{\kappa}{N}\hat{J}_{+}\hat{J}_{-},
\label{eq:heff}
\end{equation}
which generates a smooth, non-unitary time evolution between detection events. At stochastic times determined by the measurement record, the wavefunction undergoes an abrupt discontinuity corresponding to the detection of a photon emitted into the environment. Specifically, immediately after a jump the state updates as
\begin{equation}
|\psi(t_+)\rangle = 
\frac{\hat{J}_-|\psi(t_-)\rangle}
{\sqrt{\langle \psi(t_-)|\hat{J}_+\hat{J}_-|\psi(t_-)\rangle}} .
\label{eq:jump_update}
\end{equation}
The probability for such a jump to occur within a short time interval $\delta t$ is given by
\begin{equation}
\delta p
=
\delta t\,\langle \hat{L}^{\dagger}\hat{L}\rangle
=
\frac{2\kappa\,\delta t}{N}
\langle \hat{J}_{+}\hat{J}_{-}\rangle .
\end{equation}
Numerical details of the stochastic evolution follow the standard Monte Carlo Wave-Function (MCWF) method, as comprehensively reviewed in Refs.~\cite{Plenio1998_QuantumJump, Daley2014_QuantumTrajectories}.

\subsection{Entanglement Dynamics}

The entanglement properties of monitored many-body systems provide a natural probe of measurement-induced phase transitions (MIPTs). 
Precisely, one can consider the dynamics of the pure many-body states $|\psi(t)\rangle$ along its quantum trajectories and a balanced bipartition of the system into subsystems $A$ and $B$, with an equal number of atoms in the two subsystems ($N_A=N_B=N/2$).  The entanglement entropy associated with $A$ is defined as
\begin{equation}
S_A(|\psi(t)\rangle) = -\mathrm{Tr}_A \left( \hat{\rho}_A \ln \hat{\rho}_A \right),
\end{equation}
where $\hat{\rho}_A = \mathrm{Tr}_B \left[ |\psi(t)\rangle\langle\psi(t)| \right]$ denotes the reduced density matrix obtained by tracing out the degrees of freedom of subsystem $B$. 
In this way, one can estimate the steady-state entanglement by combining an ensemble average over trajectories with a long-time average over a window of duration $T_{\rm avg}$,
\begin{equation}
\overline{{S}_{N/2}}
\equiv
\frac{1}{N_{\mathrm{traj}}}
\sum_{\gamma=1}^{N_{\mathrm{traj}}}
\frac{1}{T_{\mathrm{avg}}}
\int_{t_0}^{t_0+T_{\mathrm{avg}}}
S_{N/2}^{(\gamma)}(t)\,dt ,
\label{eq:steady_entropy}
\end{equation}
where $\gamma$ labels the quantum trajectory, $N_{\mathrm{traj}}$ is the number of trajectories, and $t_0$ is chosen after the initial transient time.

Unlike linear observables (such as average spin components $\langle \hat{J}_z \rangle$), which typically relax to featureless stationary values in the ensemble average, the entanglement entropy captures the information-theoretic properties of individual quantum trajectories that are otherwise washed out by linearity.
In particular, the von Neumann entanglement entropy distinguishes regimes where quantum trajectories display extensive entanglement from those where measurements suppress correlations. Characterizing how entanglement develops along stochastic trajectories therefore provides a direct way to probe the monitored dynamics and its relation to post-selection constraints.

For the collectively driven atomic ensemble considered here, the competition between coherent driving and collective dissipation leads to different entanglement regimes controlled by the ratio $\omega_0/\kappa$.  For homogeneous initial states, the trajectory-level entanglement transition occurs at the same critical ratio as the ensemble-averaged dynamical transition, $(\omega_0/\kappa)_c=1$~\cite{Passarelli2024_ManybodyDynamics}. 
The two dynamical regimes can be distinguished by the scaling of the trajectory-averaged entanglement entropy with system size. 
For $\omega_0/\kappa<1$, the system is in a trivial stationary regime, where the half-system entanglement entropy remains weakly dependent on $N$, consistent with an area-law-like behavior. 
At the critical point, $\omega_0/\kappa=1$, the steady-state entanglement entropy grows logarithmically with system size, $S_{N/2}\sim \ln N$. 
For $\omega_0/\kappa>1$, corresponding to the boundary time-crystalline regime, the entropy growth is sub-logarithmic, $S_{N/2}\sim \ln^\beta N$, with a non-universal exponent $\beta<1$ that decreases away from the transition. 
In these regimes, the entanglement entropy grows logarithmically in time before saturation, $S(t)\sim \ln t$, and the saturation time grows only logarithmically with system size, $\tau_{\rm sat}\sim \ln N$~\cite{Passarelli2024_ManybodyDynamics}. 
This logarithmic saturation time is the key feature that reduces the post-selection overhead from exponential to polynomial scaling.

\subsection{Detector Resolution Time}

A fundamental obstacle to the experimental realization of monitored phases is the finite time resolution of the detection apparatus. Realistic detectors operate with a finite temporal resolution, $\Delta\tau$, which necessarily discretizes the continuous measurement record into finite time bins. This coarse-graining introduces an intrinsic 
ambiguity: trajectories that differ only by the occurrence or ordering of quantum jumps within a single detection window become experimentally indistinguishable. As a 
result, the conditional evolution can no longer be faithfully described by pure-state quantum trajectories, but instead must be represented by a mixed-state density 
matrix that averages over unresolved microscopic histories. Crucially, this breakdown occurs when $\Delta\tau$ becomes comparable to the average waiting 
time between emission events, $W = \langle t_{k+1} - t_k \rangle$, where $t_k$ denotes the time of the $k$-th quantum jump. When $\Delta\tau \gtrsim W$, multiple jumps within a single bin become unavoidable, leading to a degradation of state purity and a loss of trajectory-level information, obscuring the very correlations that distinguish different monitored dynamical phases.

In the context of the collective spin model introduced above, the detector-resolution problem was first analyzed in~\cite{Delmonte2024_Measurementinduced}. For homogeneous initial states, where the dynamics is confined to the 
fully symmetric Dicke sector with total spin $J = N/2$, the collective jump rate scales as $\Gamma \sim N$, so the average waiting time between quantum jumps decreases as $W \sim 1/N$. A conditional master equation accounting 
for finite-resolution effects was derived in~\cite{Delmonte2024_Measurementinduced}, showing that for bin sizes $\Delta\tau \sim \kappa^{-1}$ the entanglement transition 
can still be witnessed, but the steady-state entanglement entropy becomes sensitive to the bin size and deviates from the ideal pure-state result.

The present work addresses this subject by engineering the initial state so as to suppress the collective jump rate and increase $W$, thereby preserving the visibility of trajectory-level measurement-induced phenomena under realistic experimental constraints.

\begin{figure}[h]
    \centering
    \includegraphics[scale=0.58]{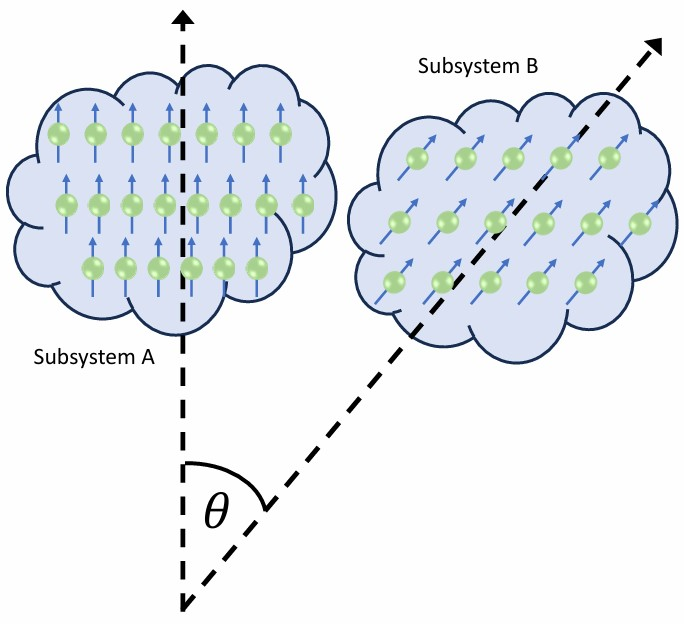}
    \caption{Schematic representation of the inhomogeneous initial-state preparation. The system is divided into two internally permutation-symmetric collective spin ensembles, $A$ and $B$. The ensembles are $\theta$ rotated from each other - as given by Eq.\eqref{eq:initial_state} -  with this angle thus controlling the degree of inhomogeneity between them.}
    \label{fig:rotation}
\end{figure}

\section{Inhomogeneous Initial State Preparation}
\label{inhom}
A key element of our work is the preparation of the system in an initial state with a controllable degree of spatial inhomogeneity. We consider the system as two collective spin ensembles, $A$ and $B$, each containing $N/2$ atoms - see Fig.\ref{fig:rotation}. Given the ensembles are not initially aligned, such an inhomogeneous state breaks the global permutation symmetry of the full $N$-spin ensemble, while preserving permutation symmetry within each of the two subsystems. The corresponding collective spin operators are defined as
\begin{equation}
\hat{J}^{A}_{\alpha}
=
\frac{1}{2}\sum_{i\in A}\hat{\sigma}^{\alpha}_{i},
\qquad
\hat{J}^{B}_{\alpha}
=
\frac{1}{2}\sum_{i\in B}\hat{\sigma}^{\alpha}_{i},
\qquad
\alpha=x,y,z .
\end{equation}


Since each subsystem remains internally permutation symmetric, the Hilbert space of each ensemble is spanned by Dicke states $|J_{\eta},m_{\eta}\rangle$, with $\eta=A,B$, satisfying
\begin{eqnarray}
    (\hat{\mathbf{J}}^{\eta})^2 |J_{\eta},m_{\eta}\rangle
&=&
J_{\eta}(J_{\eta}+1)|J_{\eta},m_{\eta}\rangle, \\
\hat{J}^{\eta}_{z}|J_{\eta},m_{\eta}\rangle
&=&
m_{\eta}|J_{\eta},m_{\eta}\rangle .
\end{eqnarray}
with,
\begin{equation}
J_A=J_B=\frac{N}{4},
\qquad
m_{\eta}=-J_{\eta},-J_{\eta}+1,\ldots,J_{\eta}.
\end{equation}


The full many-body state is then expanded in the tensor-product Dicke basis as
\begin{equation}
|\psi(t)\rangle
=
\sum_{m_A=-J_A}^{J_A}
\sum_{m_B=-J_B}^{J_B}
c_{m_A,m_B}(t)
\,|J_A,m_A\rangle\otimes |J_B,m_B\rangle .
\label{eq:two_ensemble_state}
\end{equation}
This representation reduces the effective Hilbert-space dimension from $2^N$ to
\begin{equation}
d=(2J_A+1)(2J_B+1)
=
\left(\frac{N}{2}+1\right)^2,
\end{equation}
while retaining the bipartite structure required to describe the inhomogeneous initial states.

The protocol begins with all spins in the fully polarized state along the z-axis, $|\Uparrow\rangle = |\uparrow\rangle_1 \otimes \dots \otimes |\uparrow\rangle_N$. We then apply a rotation around the x-axis to the spins in subsystem B, as shown in Fig.~\ref{fig:rotation}:
\begin{equation}
    |\psi_\theta(t_0)\rangle = \exp(-i\theta \hat{J}_x^B) |\Uparrow\rangle,
    \label{eq:initial_state}
\end{equation}
The angle $\theta$ serves as a continuous control parameter for the relative orientation, and hence the inhomogeneity, between the two collective ensembles. 

Both ingredients of this protocol are within reach of current experiments. A natural platform is cavity QED with atomic ensembles, where collective decay of the form $\hat{L}=\sqrt{2\kappa/N}\,\hat{J}_-$ arises when all atoms couple identically to a shared, strongly damped cavity mode. Full polarization along $z$ is routinely achieved by optical pumping~\cite{Deutsch2010}. Separate atomic ensembles can be positioned inside an optical cavity with optical tweezers and imaged individually through the cavity emission, as demonstrated with up to $25$ tweezer-trapped ensembles in confocal multimode cavity QED~\cite{Kroeze2023,Marsh2025}. The rotation of one sub-ensemble can then be implemented by local addressing: high-fidelity local rotations of tweezer-trapped atoms through controlled atom movements~\cite{Shaw2024}, and site-selective control in large trapped-ion crystals~\cite{Makadia2026}, have recently been demonstrated. For our protocol the two ensembles must couple to the cavity mode with equal strength, so that the decay remains collective over the full system. The balanced condition $N_A=N_B$ can in turn be enforced by the trap geometry or by post-selecting on the atom number.

\section{Results}
\label{sec:results}

In this section, we study the effects of the inhomogeneous initial-state preparation on the quantum trajectories of the collective spin model, focusing on 
two key aspects: the detector-resolution problem and the corresponding entanglement dynamics. We begin in Sec.~\ref{sec:waiting} by providing analytical and physical insights behind the mitigation of the detector-resolution barrier, as well as numerical simulations corroborating  these expectations.  
In Sec.~\ref{sec:entanglement}, we then characterize the entanglement dynamics along the quantum trajectories, highlighting the influence on the inhomogeneities over the saturation time and steady state properties.

All our simulations are performed by numerically integrating the stochastic Schrödinger equation derived from the quantum jump method. We integrate in first order $O(dt)$, using a time step of $\kappa dt = 0.01/N$. 
To obtain statistically robust results for observables such as entanglement entropy and waiting times, we average all computed quantities over an ensemble of $250$ independent quantum trajectories for each parameter set. For the waiting time in particular, we collect all inter-jump intervals recorded along a given trajectory over the full simulation window $\kappa t\in[0,200]$ and take their mean as the trajectory-level estimate of $W$; the reported $W$ is the average of these trajectory-level estimates over the $250$ trajectories.

\subsection{Waiting Time Statistics}
\label{sec:waiting}

To quantify the mitigation of the detector resolution problem, we analyze the average waiting time $W$, defined as the mean duration between consecutive quantum jump events along a single trajectory. 
First, we provide analytical arguments in the form of a bound for the effects of inhomogeneities on the waiting time and its potential mitigation of the detector resolution time.

\textit{Analytical bound.- } Due to total angular momentum 
conservation, the weight of the state in 
each $J$ sector is conserved on average over the trajectories. 
This implies that the average jump rate over the trajectories,
\begin{eqnarray}
    \Gamma(t) &=& \sum_{j=1}^{N_{\textrm{traj}}} 
    \frac{ \langle \hat{L}^\dagger \hat{L} \rangle_t}{N_{\textrm{traj}} } \nonumber  = 
\frac{2\kappa}{N} \sum_{j=1}^{N_{\textrm{traj}}} \frac{\langle \psi(t)| \hat{J}_+\hat{J}_- |\psi(t)\rangle}{N_{\textrm{traj}}},
\end{eqnarray} 
and recalling that $\langle \hat{J}_+\hat{J}_- \rangle
=
\langle \hat{J}^2 \rangle
-
\langle \hat{J}_z^2 \rangle
+
\langle \hat{J}_z \rangle$, it will be  largely controlled by the population of the $J$-sectors settled
by the initial condition - in other words, by the initial 
inhomogeneities in the system preparation. Noticing that one 
could roughly relate the average waiting time to the average 
decay rate as $W(t) \sim 1/\Gamma(t)$, we thus see how 
the inhomogeneities are directly related to the waiting time 
and the detector resolution problem.

More precisely, we can use the angular-momentum identity,
$ \langle \hat{J}_+\hat{J}_- \rangle
=
\langle \hat{J}^2 \rangle
-
\langle \hat{J}_z^2 \rangle
+
\langle \hat{J}_z \rangle,$
and since $\langle \hat{J}_z^2 \rangle \ge 0$, one obtains 
the upper bound,
$\langle \hat{J}_+\hat{J}_- \rangle \le \langle \hat{J}^2 \rangle
+ \langle \hat{J}_z \rangle.$ 
The leading term in this inequality is in general  given by the total angular momentum, scaling quadratically with the number of spins ($O(N^2)$) on top of a linear scaling for the magnetization. In this way we may further approximate it to 
$\langle \hat{J}_+\hat{J}_- \rangle \lesssim \langle \hat{J}^2 \rangle $. Recalling that the angular momentum is conserved
on average over the trajectories, this 
 implies a lower bound on the average waiting time through all evolution,
\begin{equation}
    W(t) \gtrsim  \frac{N}{2\kappa \langle \hat{J}^2 \rangle_{t=0}},
\label{eq:W_bound}
\end{equation}
where $\langle \hat{J}^2 \rangle_{t=0}$ is  determined by the initial state.

\begin{figure}
    \centering
    \includegraphics[scale=0.35]{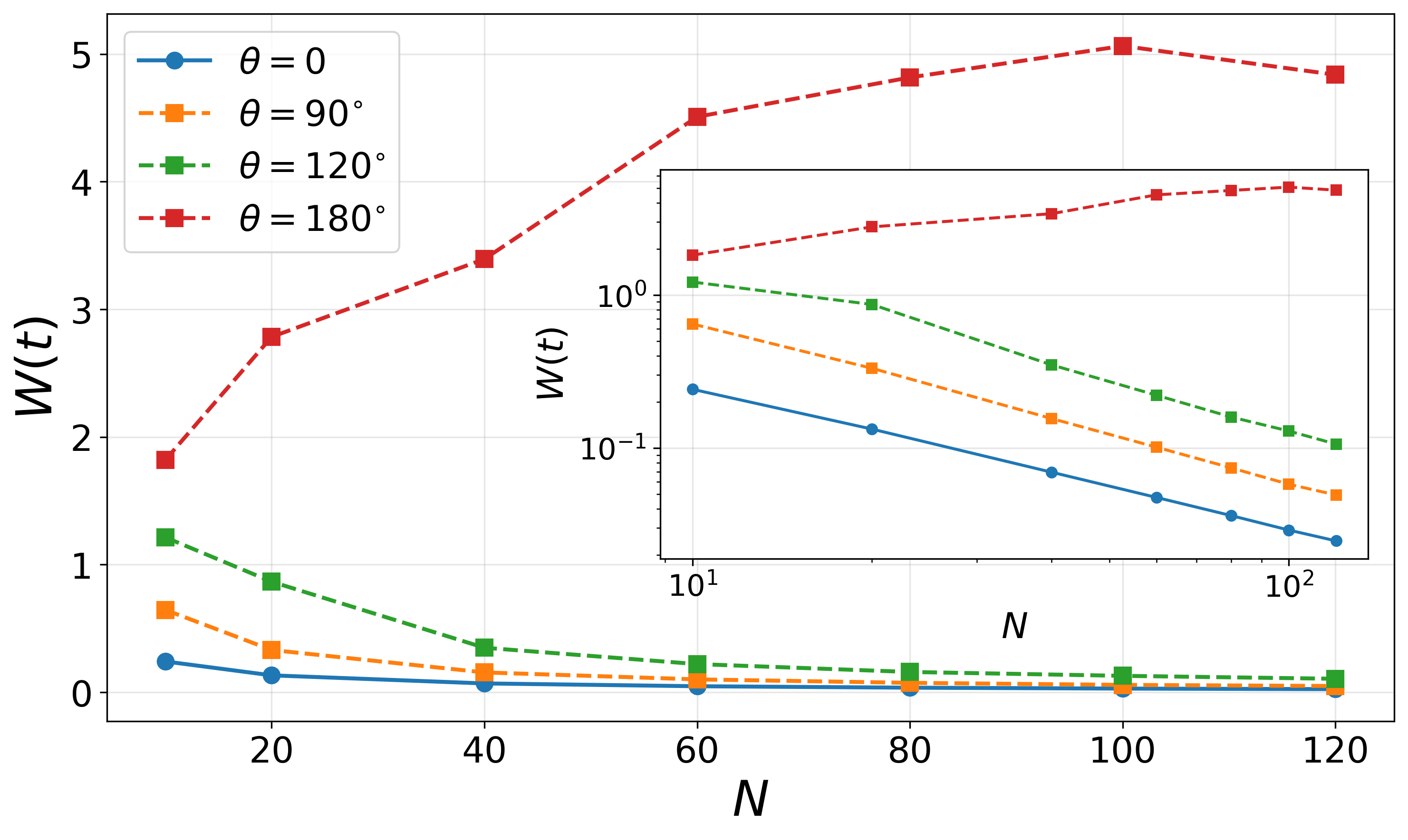}
    \caption{Average waiting time $W(t)$ versus system size $N$ for different initialization angles $\theta$.  For the anti-aligned case ($\theta=180^\circ$), collective decay is suppressed and $W(t)$ remains finite of $\mathcal{O}(1)$, while for intermediate angles it shows a waiting time decreasing with $N$. \textit{Inset}: same plot in log-log scale, showing clearly an algebraic decay $W(t) \sim N^{-1}$ for the homogeneous and intermediate angle cases. 
    }
    \label{fig:waiting_time}
\end{figure}

\begin{figure}
    \centering
    \includegraphics[width=0.50\textwidth]{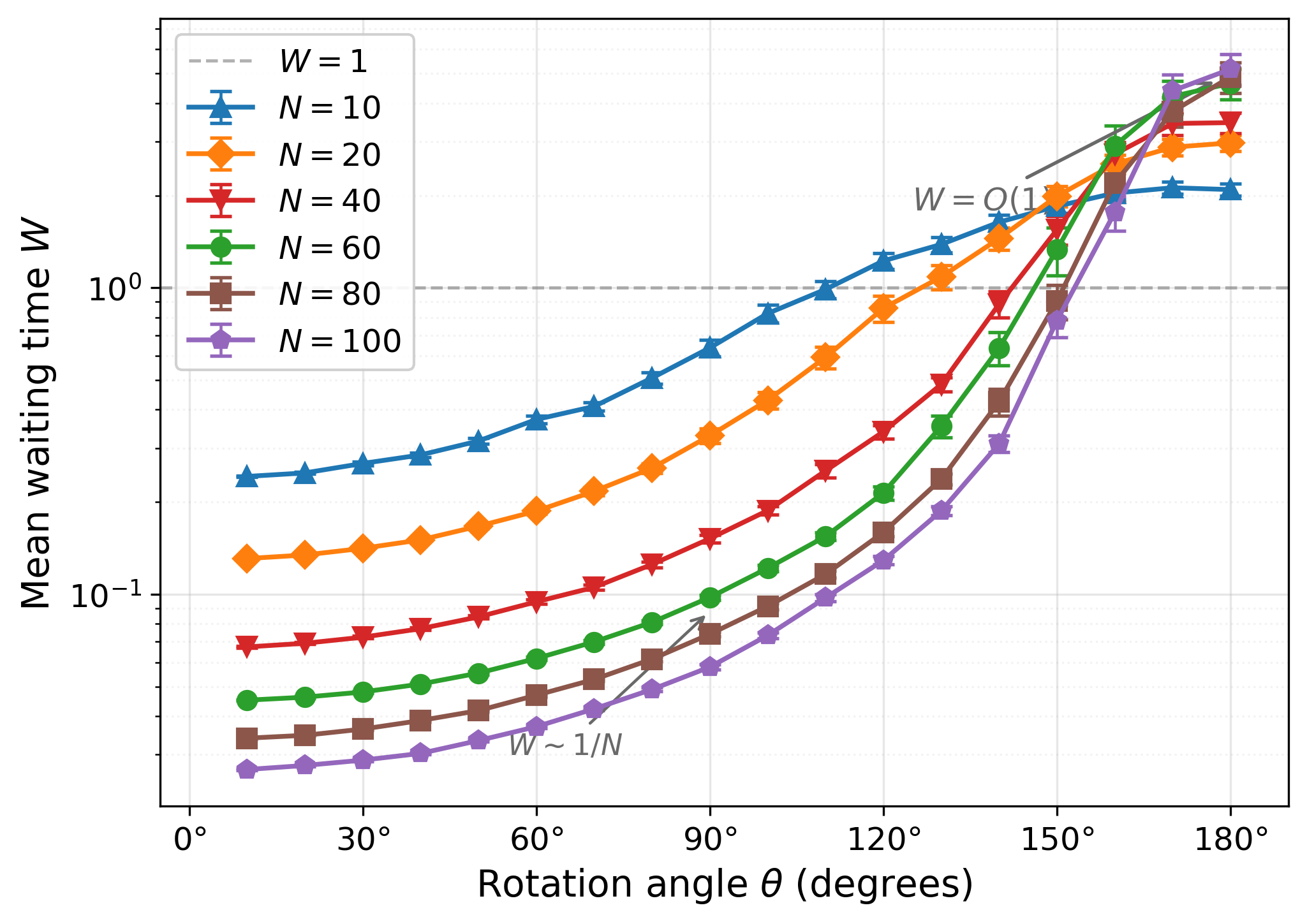}
    \caption{Average waiting time $W$ as a function of the inhomogeneity angle $\theta$, at fixed $\omega_0/\kappa=2.0$ for $N=10$, $20$, $40$, $60$, $80$, and $100$, on a logarithmic vertical scale. Error bars indicate the standard error on the mean over $250$ trajectories and are mostly smaller than the marker size. The waiting time grows monotonically with $\theta$ over the whole range, spanning nearly two orders of magnitude between the homogeneous and the anti-aligned configurations. The dashed line marks $W=1$.}
    \label{fig:waiting_time_theta}
\end{figure}

For a general inhomogeneity we can further expand the total spin operator in terms of the two subsystems  A and B contributions:
\begin{align}
    \langle \hat{{J}}^2 \rangle &= \langle (\hat{{J}}_A + \hat{ {J}}_B)^2 \rangle \nonumber \\
    &= \underbrace{\langle \hat{ {J}}_A^2 \rangle + \langle \hat{ {J}}_B^2 \rangle}_{\text{Local Terms}} + \underbrace{2 \langle \hat{ {J}}_A \cdot \hat{ {J}}_B \rangle}_{\text{Interference Term}}.
    \label{eq:J2_expansion}
\end{align}
On the one hand, the "Local Terms" are fixed by the subsystem sizes ($J_{A/B} = N/4$):
\begin{equation}
    \langle \hat{ {J}}_A^2 \rangle + \langle \hat{ {J}}_B^2 \rangle = 2 \times \frac{N}{4}\left(\frac{N}{4} + 1\right) 
\end{equation}

The ``Interference Term'' on the other hand is dependent 
on the initial state. For the initial state of Eq.\eqref{eq:initial_state} the
subsystem $A$ remains fully polarized so $\langle\hat{J}_\pm^A\rangle = 0$, 
and only the longitudinal part of the inner product contributes:
\begin{equation}
2\langle \hat{J}_A \cdot \hat{J}_B \rangle = 
2\langle \hat{J}_z^A \rangle\langle \hat{J}_z^B \rangle = 
\frac{N^2}{8}\cos\theta,
\end{equation}
where we used $\langle \hat{J}_z^A \rangle = N/4$ and 
$\langle \hat{J}_z^B \rangle = (N/4)\cos\theta$.
Substituting these back into Eq.~(\ref{eq:J2_expansion}) 
we obtain that,
\begin{align}
\langle \hat{J}^2 \rangle 
&= \frac{N^2}{8}(1+\cos\theta) + \frac{N}{2}.
\end{align}

In this way, the waiting time bound for a general inhomogeneity angle $\theta$ is:
\begin{equation}
W \gtrsim \frac{N}{2\kappa  \left( \frac{N^2}{8}(1+\cos\theta) + \frac{N}{2} \right)}
\label{eq:W_bound_general}
\end{equation}
This bound has different behaviors for large ensembles:
\begin{itemize}
\item  \textit{Homogeneous case} ($\theta=0$): here $W \gtrsim O(1/N)$, and the detector-resolution constraint as determined by the bound can become increasingly severe in the thermodynamic limit.

\item \textit{Intermediate angles} ($0 < \theta < \pi$): 
the waiting time still decays extensively with $N$, $W \gtrsim O(1/N)$,  albeit with a prefactor $(1+\cos\theta)^{-1}$ that depends on the initial state's relative orientation. Thus, apart from the detrimental resolution time with system size, it would still allow for tunability via initial state preparation.

\item \textit{Anti-aligned case} ($\theta=\pi$):  the interference terms exactly cancel the local ones at leading order, and the bound simplifies to,
\begin{equation}
W \gtrsim \frac{1}{\kappa} \sim \mathcal{O}(1).
\end{equation}
i.e., it remains \textit{finite and independent 
of system size}. 
\end{itemize}

\textit{Numerical simulations.- } To corroborate these analytical insights, we numerically compute the waiting time for varying inhomogeneities. Our results are presented in Fig.~\ref{fig:waiting_time}. Indeed, homogeneous ensembles exhibit waiting times that vanish with system size, following an algebraic decay $\sim \mathcal{O}(1/N)$. For intermediate angles $0<\theta<\pi$, this algebraic decay persists, but with an overall multiplicative factor that increases as the system approaches the anti-aligned configuration. In the extreme anti-aligned case, the waiting time remains finite in the large-$N$ limit, thereby fully mitigating the detector-resolution issue.

Since Fig.~\ref{fig:waiting_time} shows $W$ against $N$ at fixed $\theta$, the enhancement at intermediate angles appears only as a change of prefactor and is hard to read off directly. We therefore show in Fig.~\ref{fig:waiting_time_theta} the complementary cut, $W$ as a function of $\theta$ at fixed $N$. The waiting time grows monotonically over the whole range, and all curves remain separated at intermediate angles, confirming that the $\mathcal{O}(1/N)$ scaling is preserved while the prefactor is enhanced monotonically with $\theta$. We note that the system-size-independent regime $W\sim\mathcal{O}(1)$ is reached strictly only at the exact anti-aligned point, and for $\theta<\pi$ the waiting time still scales as $1/N$: the protocol does not eliminate this scaling, but for the system sizes considered here a deviation of $10^\circ$--$20^\circ$ from $\theta=\pi$ already yields $W$ of order a few $\kappa^{-1}$, comparable in practice to the ideal anti-aligned value.

\subsection{Entanglement Entropy}
\label{sec:entanglement}

\begin{figure} [h]
    \includegraphics[width=0.49 \textwidth]{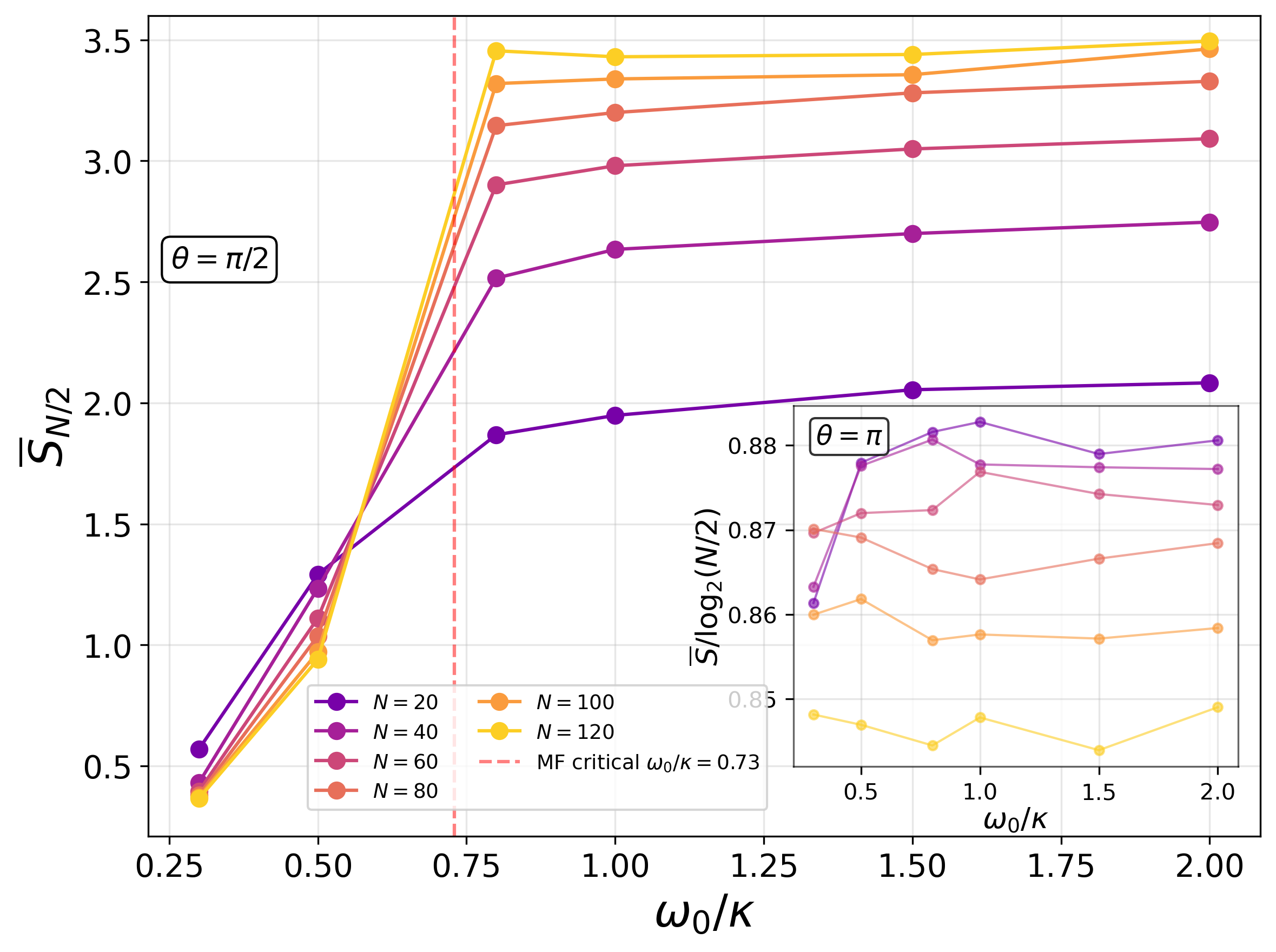}
    \caption{ Steady-state entanglement entropy for the inhomogeneous case $\theta=\pi/2$, shown across the phase diagram. Two distinct regimes are visible: for low couplings, the entanglement follows an area-law scaling, while for higher couplings a sub-logarithmic scaling emerges. The dashed vertical line indicates the mean-field critical coupling obtained from magnetization observables. The inset displays the anti-aligned configuration ($\theta=\pi$), where the entanglement entropy is rescaled by $\log_2 (N/2)$ indicating its logarithmic growth with system size.
    } 
    \label{fig.phase.diagram.ent}
\end{figure}

\begin{figure*}
    \includegraphics[scale=0.0955]{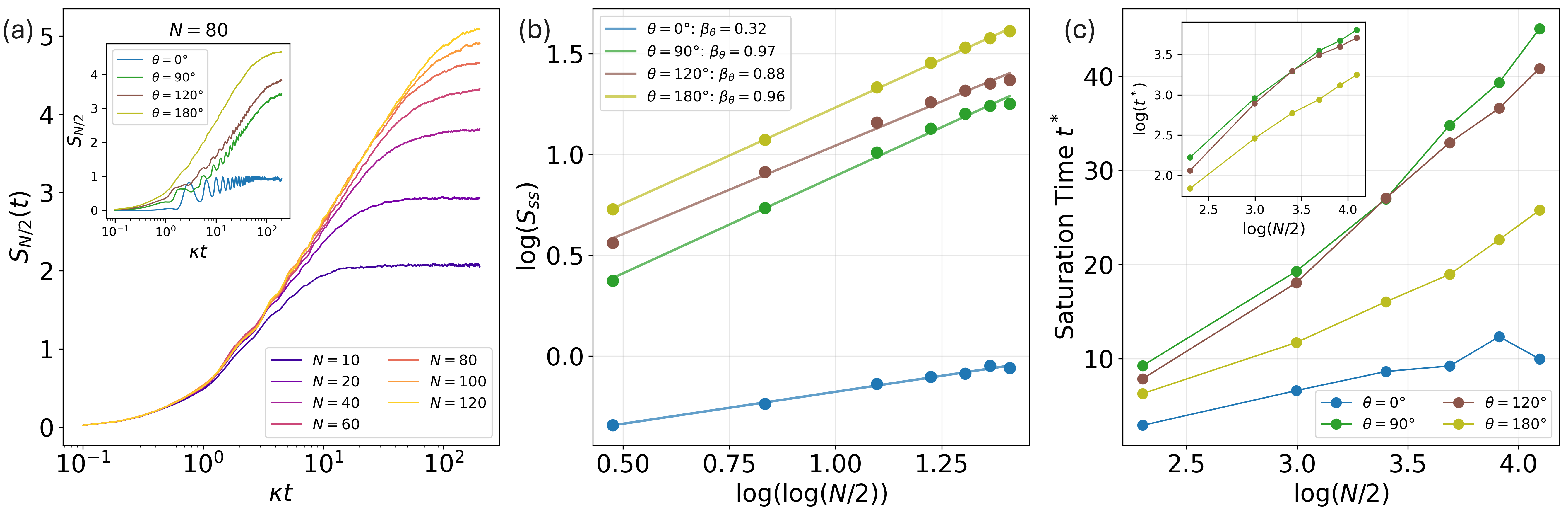}
    \caption{  \textbf{Entanglement and inhomogeneities in the BTC phase ($\omega_0/\kappa=2$). } \textbf{(a)} Entanglement dynamics for $\theta=\pi$ and different system sizes. The time axis is rescaled to highlight the logarithmic growth in time observed in the  dynamics. The (inset panel) shows the dynamics for various inhomogeneity angles $\theta$ at fixed system size $N=80$. \textbf{(b)} Steady-state entanglement entropy for different inhomogeneity angles and system sizes. The axes are rescaled to highlight a sub-logarithmic scaling $\bar{S}_{N/2} \sim \ln^{\beta_\theta}(N/2)$, with fitted exponent $\beta_\theta$ depending on the inhomogeneity. \textbf{(c)} Saturation time $t^*$ as a function of system size for various inhomogeneity angles. The inhomogeneous cases ($\theta \neq0$) show a saturation time larger than logarithmically in $N$, with a (inset panel) tendency towards polynomial scaling $t^* \sim N^{\alpha_\theta}$ - though we note that the fit is not fully conclusive for the largest system sizes.
    The saturation time is extracted by fitting the trajectory-averaged entropy to $\bar{S}_{N/2}(t)=a+b\,e^{-\kappa t/t^*}$, following the procedure of Ref.~\cite{Passarelli2024_ManybodyDynamics} (see Appendix~\ref{app:saturation}), so that $t^*$ is the $e$-folding time of the approach to the asymptotic value $a$ rather than the time at which saturation is visually complete.}
    \label{fig.ent.dynamics.saturation}
\end{figure*}

In this section, we explore how inhomogeneities affect the entanglement properties of the system, with particular attention to their implications for the postselection barrier. Our results suggest a nuanced picture in which the overall structure remains stable, while the dynamical behaviour - especially the saturation times - exhibits a nontrivial dependence on the inhomogeneity degree.

We first examine whether the MIPT survives the introduction of inhomogeneities. As shown in Fig.~\ref{fig.phase.diagram.ent}, the steady-state entanglement entropy clearly indicates the presence of two distinct regimes, separated by a critical coupling that agrees well with the mean-field prediction for the stationary-to-BTC transition. In the low drive-to-dissipation regime, the entanglement follows an area-law scaling, characteristic of a stationary phase with limited correlations. For higher ratios, within the BTC region, we observe a transition to a phase where entanglement grows with system size, following a sub-logarithmic scaling $\bar{S}_{N/2}\sim\ln^{\beta_\theta}(N/2)$. A particularly interesting case is the anti-aligned configuration, where the system remains in a BTC state for any finite coupling. In this case, we find a logarithmic growth of entanglement with system size across all parameters (see inset of Fig.~\ref{fig.phase.diagram.ent}). It is worth noting that such logarithmic scaling is observed at the critical transition in homogeneous systems \cite{Fazio2025_ManyBodyOpen}, not strictly required in the inhomogeneous case.

We now turn our attention to the BTC phase, where the postselection-free barrier plays a central role. In Fig.~\ref{fig.ent.dynamics.saturation} we present a detailed analysis of the entanglement dynamics, including its growth, saturation time, and steady-state values. As shown in Fig.~\ref{fig.ent.dynamics.saturation}(a), for intermediate times (before saturation sets in) the entanglement entropy grows logarithmically in time for all inhomogeneity angles $\theta$. This behaviour closely resembles that of the homogeneous case, indicating that the intermediate-time dynamics are qualitatively insensitive to the inhomogeneities. However, the later stages, particularly the saturation values and the time required to reach them, are nontrivially influenced by $\theta$.

Concerning the steady-state entanglement entropy, Fig.~\ref{fig.ent.dynamics.saturation}(b) indicates that for all angles the entropy remains macroscopically large, increasing  with system size. The steady-state entanglement is described by a sub-logarithmic form $\ln^{\beta_\theta}(N/2)$, where the exponent $\beta_\theta$ depends on the inhomogeneity.

With regard to saturation time, we recall that, in the homogeneous limit, it is known to scale logarithmically with $N$, a key factor in the postselection-free property. In the presence of inhomogeneities, however, our finite-size data (Fig.~\ref{fig.ent.dynamics.saturation}(c)) show a tendency towards a slower growth. For the system sizes accessible to us, a polynomial fit $t^* \sim N^{\alpha_\theta}$, with $\alpha_\theta >0$, provides a reasonable description, with intermediate angles $\theta$ yielding the longest times. Nevertheless, we note that the fit is not fully satisfactory for all data points. In particular, for the largest system sizes - especially for $\theta = \pi/2$ and $2\pi/3$ - there is a visible tendency towards a slower increase, which may indicate that the asymptotic scaling could be less severe than a pure polynomial. Given the limitations imposed by our finite-system simulations, we cannot draw a definitive conclusion regarding the exact functional form of the scaling. We can only state that the observed trend is a departure from the logarithmic behaviour of the homogeneous case, and that this departure merits further investigation with larger system sizes.

This point is particularly important in light of the postselection-free aspect of the model. If the saturation time were to scale more rapidly than logarithmically with $N$, the system would lose its advantage of fast saturation, thereby reintroducing the postselection barrier (at least partially, depending on how large is the saturating time scaling) that the model was originally designed to circumvent. Thus, while inhomogeneities may offer practical advantages - for instance, by improving the resolution of the detection of the phase transition - they also come with a potential cost: the need to wait longer for the system to reach its steady state. The trade-off between these competing factors is likely to depend on the specific value of $\theta$, and a careful optimisation may be required depending on the experimental context.

\section{Conclusion}
\label{sec:Conclusion}

In this work, we have investigated how initial-state inhomogeneities affect the waiting time statistics and entanglement dynamics of a collectively monitored spin model exhibiting a boundary time-crystalline phase. Our analysis reveals a nuanced picture in which inhomogeneities provide significant control over the waiting time statistics (and therefore for the detector's resolution problem), albeit at the cost of modifying the postselection-free character of the model. Specifically, we first showed that the MIPT survives under inhomogeneities, with the steady-state entanglement displaying area-law and sub-logarithmic phases. The average waiting time between quantum jumps increases significantly with $\theta$, decreasing with $N$ for $\theta \neq \pi$ but with a strongly enhanced prefactor, while for $\theta = \pi$ it remains finite, fully mitigating the detector-resolution barrier. This mitigation, however, comes at a cost: the entanglement saturation time, which in the homogeneous case grows logarithmically with $N$, tends towards slower growth in the presence of inhomogeneities - with our finite-size data suggesting a polynomial scaling -  thereby partially reintroducing the postselection barrier.

 Beyond these central findings, our work suggests several interesting avenues for further investigation. Firstly, the observed deviations from the homogeneous behaviour, especially regarding saturation times, raise interesting questions about the robustness of the postselection-free barrier in more general settings, such as spin-boson models with BTC behavior \cite{PhysRevA.108.062216,PhysRevLett.130.180401}. Moreover, whether the scaling remains polynomial, crosses over to a different functional form, or recovers logarithmic behaviour at larger system sizes remains an open question that would benefit from future studies on larger systems and possibly from analytical approaches as spin-wave theory \cite{Li2024_MonitoredLongRange}. Secondly, it would be interesting to explore whether similar waiting-time control could be achieved through other mechanisms, such as different interacting drives or engineered dissipation, or through jump operators that do not share the strong symmetry $[\hat{J}^2,\hat{L}]=0$ exploited here, for which the enhancement mechanism could be different.  Thirdly, our findings demonstrate that initial-state inhomogeneities can be exploited to engineer highly entangled steady states, offering an intriguing approach to generating large-scale entanglement in dissipative systems that do not rely on usual approaches such as dark state engineering \cite{PhysRevA.78.042307}, and connecting to recent work exploiting initial-state control more broadly to select among multistable entangled steady states in collective spin ensembles \cite{Padilla2025_GeneratingEntangled}.

\begin{acknowledgments}
 We acknowledge financial support from the Brazilian funding agencies CAPES, CNPq (308637/2022-4), FAPERJ (No. E-26/210.236/2024, and No.E-26/204.340/2025), and by the Serrapilheira Institute (grant number Serra 2211-42166).
\end{acknowledgments}

\bibliographystyle{modified-apsrev4-2}
\bibliography{references}

\clearpage

\onecolumngrid
\appendix

\section{Mean-Field Analysis of the BTC-to-Stationary Transition}
\label{app:meanfield}

\begin{figure}[!ht]
\centering
\begin{minipage}[t]{0.48\linewidth}
    \centering
    \textbf{(a)}\\[2pt]
    \includegraphics[width=\linewidth]{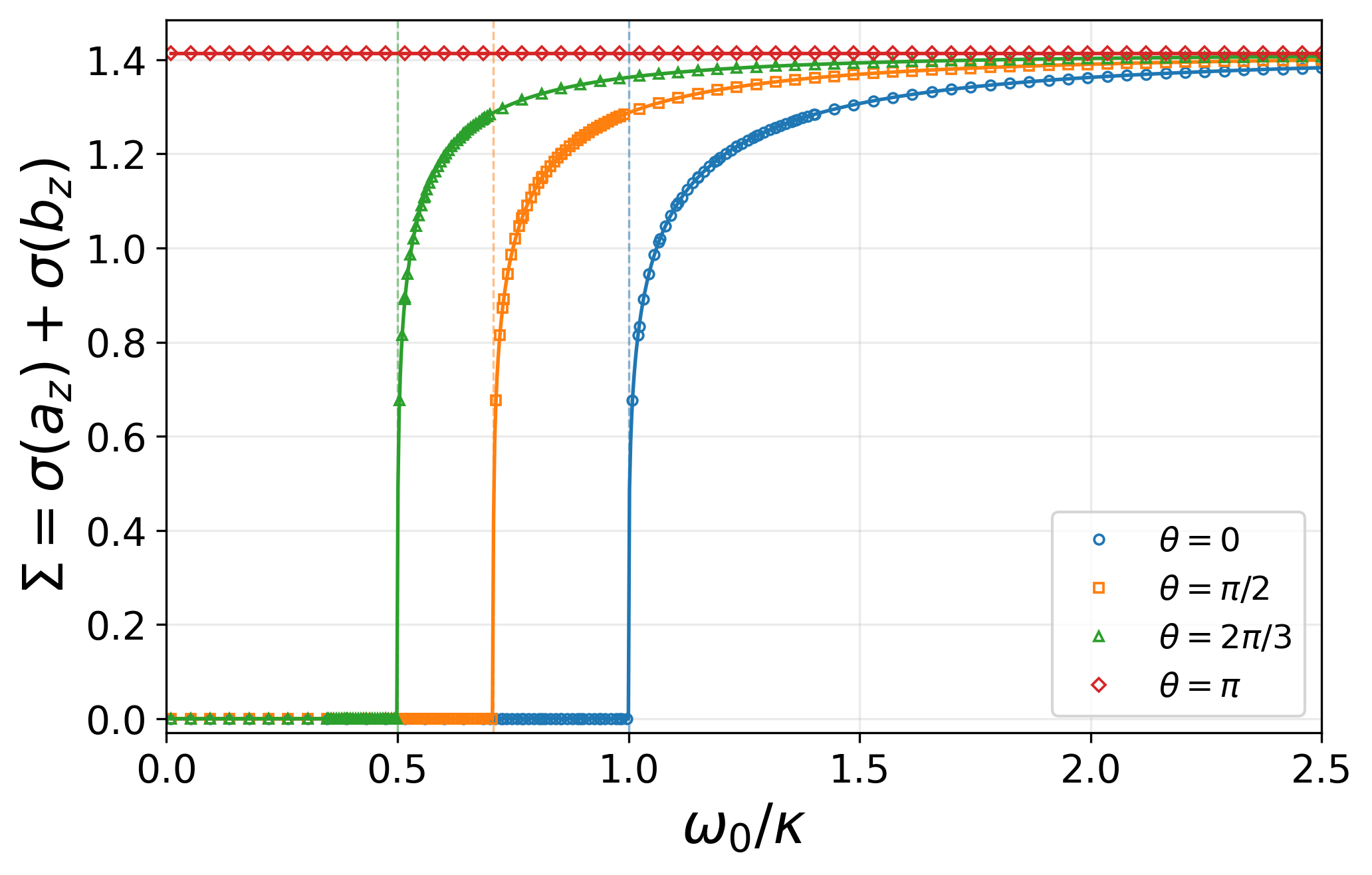}
\end{minipage}\hfill
\begin{minipage}[t]{0.48\linewidth}
    \centering
    \textbf{(b)}\\[2pt]
    \includegraphics[width=\linewidth]{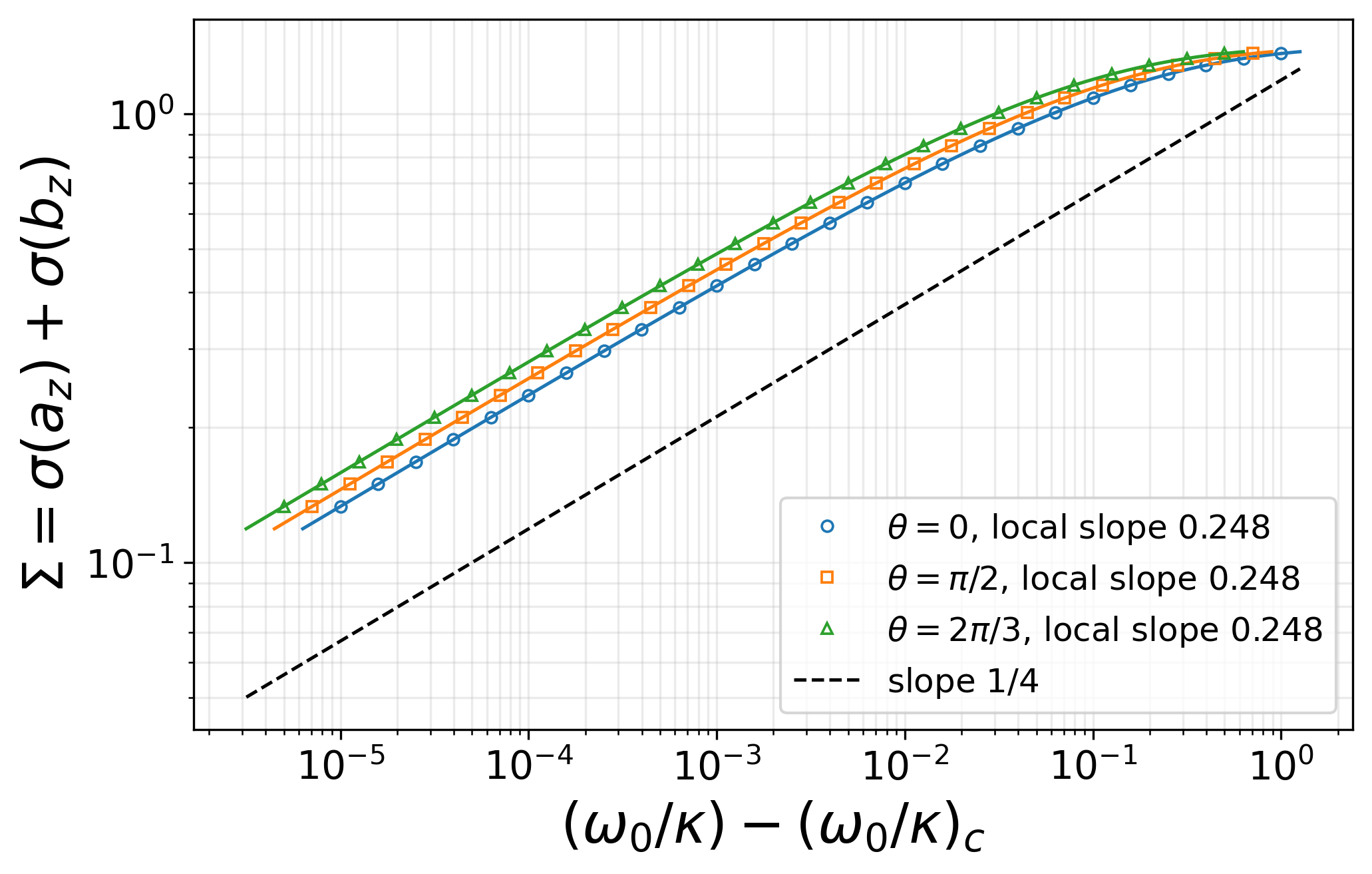}
\end{minipage}
\caption{Mean-field order parameter $\Sigma=\sigma_t(a_z)+\sigma_t(b_z)$
across the BTC-to-stationary transition.
(a) $\Sigma$ as a function of $\omega_0/\kappa$ for four values of
$\theta$. Symbols represent numerical integration of the mean-field
equations~\eqref{eq:mf_eom}; the analytical critical point
Eq.~\eqref{eq:critical_point} is marked by vertical dashed lines. For
$\theta=\pi$, $\Sigma=\sqrt{2}$ for all $\omega_0>0$.
(b) Log-log plot of $\Sigma$ versus the distance from the critical
point, showing the critical scaling
$\Sigma\propto[(\omega_0/\kappa)-(\omega_0/\kappa)_c]^{1/4}$
(dashed line) for all $\theta<\pi$.}
\label{fig:mf_op}
\end{figure}

\twocolumngrid

\subsection{Mean-field equations of motion}
In the thermodynamic limit the dynamics of the two-sub-ensemble model
admits an exact mean-field description~\cite{Carollo_2021}. Defining the
normalized sub-ensemble magnetizations $\vec{a}$ and $\vec{b}$ through
their components
\begin{equation}
a_\alpha=\frac{4\langle\hat{J}^A_\alpha\rangle}{N},\qquad
b_\alpha=\frac{4\langle\hat{J}^B_\alpha\rangle}{N},\qquad
\alpha=x,y,z,
\end{equation}
and their mean $\vec{m}=(\vec{a}+\vec{b})/2$,
the Lindblad master equation~\eqref{eq:lindblad_general} reduces under
mean-field factorization to the coupled equations of motion
\begin{equation}
\dot{\vec{a}}=\vec{a}\times\vec{\Omega},\qquad
\dot{\vec{b}}=\vec{b}\times\vec{\Omega},
\label{eq:mf_eom}
\end{equation}
where the effective precession field is
\begin{equation}
\vec{\Omega}=-\omega_0\hat{x}+\kappa\,\vec{m}\times\hat{z}.
\label{eq:Omega}
\end{equation}
The Hamiltonian $\hat{H}=\omega_0\hat{J}_x$ drives precession about the
$x$-axis, while the jump operator $\hat{L}=\sqrt{2\kappa/N}\,\hat{J}_-$
contributes the dissipative term through $\vec{m}\times\hat{z}$.

\subsection{Analytical critical point and oscillation frequency}

A key structural property of Eqs.~\eqref{eq:mf_eom} is that the relative
angle between $\vec{a}$ and $\vec{b}$, and hence the magnitude
$|\vec{m}|=\cos(\theta/2)$, is conserved throughout the dynamics. This
allows one to reduce the two-sub-ensemble problem to an effective
single-vector problem with a renormalized coupling
\begin{equation}
\kappa_{\mathrm{eff}}=\kappa\cos\!\left(\frac{\theta}{2}\right).
\label{eq:kappa_eff}
\end{equation}
The critical point of this effective model is $\omega_0=\kappa_{\rm eff}$,
which gives the analytical result
\begin{equation}
\left(\frac{\omega_0}{\kappa}\right)_c(\theta)=\cos\!\left(\frac{\theta}{2}\right),
\qquad 0\leq\theta<\pi.
\label{eq:critical_point}
\end{equation}
Below this threshold the magnetizations relax to a stationary fixed point;
above it they rotate periodically with period
\begin{equation}
T=\frac{2\pi}{\sqrt{\omega_0^2-\kappa_{\rm eff}^2}}.
\end{equation}
Setting $r=\omega_0/\kappa-(\omega_0/\kappa)_c$ as the distance above the
critical point, the oscillation frequency approaches zero as
\begin{equation}
\omega_{\rm osc}=\frac{2\pi}{T}\propto r^{1/2},\quad r\to 0^+,
\end{equation}
so the period diverges and the transition is continuous. The symbols in
Fig.~\ref{fig:mf_phase_diagram} represent a numerical determination of
the phase boundary from the mean-field equations and serve as a check of
Eq.~\eqref{eq:critical_point}.

At $\theta=\pi$ one has $|\vec{m}|=0$, so $\vec{\Omega}=-\omega_0\hat{x}$
and both sub-ensembles precess freely. The phase boundary reaches
$(\omega_0/\kappa)_c=0$, meaning the BTC phase extends to arbitrarily weak
drive.

\subsection{Mean-field order parameter and critical exponent}

To characterize the transition quantitatively we introduce the order
parameter
\begin{equation}
\Sigma=\sigma_t(a_z)+\sigma_t(b_z),
\label{eq:Sigma}
\end{equation}
where $\sigma_t(a_z)$ denotes the temporal standard deviation of $a_z(t)$
computed along the long-time mean-field trajectory,
\begin{equation}
\sigma_t(a_z)=\left[\,\overline{a_z^2}-\overline{a_z}^{\,2}\,\right]^{1/2},
\end{equation}
with the time average
\begin{equation}
    \overline{a_z}=\frac{1}{\tau}\int_{t_0}^{t_0+\tau}a_z(t)\,dt,
\end{equation}
and $\overline{a_z^2}$ defined in the same way. This quantity vanishes
when the magnetization relaxes to a fixed point and remains finite for
persistent oscillations, providing a direct measure of the temporal
variation that distinguishes the two phases. It is distinct from the
standard deviation across quantum trajectories at a fixed time, which
measures trajectory-to-trajectory fluctuations; those two quantities are
compared explicitly in Fig.~\ref{fig:mf_phase_diagram_1}.

As shown in Fig.~\ref{fig:mf_op}, $\Sigma$ is finite in the BTC phase
and vanishes continuously at the critical point. Numerically, the
log-log plot of $\Sigma$ versus the distance from the critical point
yields a local slope of approximately $0.248$ in the range
$10^{-5}\leq c-1\leq 10^{-4}$, where
$c=(\omega_0/\kappa)/(\omega_0/\kappa)_c$. This is consistent with the
mean-field critical exponent
\begin{equation}
\Sigma\propto
\left[\frac{\omega_0}{\kappa}
-\left(\frac{\omega_0}{\kappa}\right)_c\right]^{1/4},
\end{equation}
which holds for all $0\leq\theta<\pi$. The transition is therefore
continuous with $\beta=1/4$ at the mean-field level; establishing the
universality class beyond mean-field theory would require further
analysis. For $\theta=\pi$ the sub-ensembles precess freely and
$\Sigma=\sqrt{2}$ for any $\omega_0>0$.

\onecolumngrid

\section{Extraction of the Saturation Time}
\label{app:saturation}

\begin{figure}[h!]
\centering
\includegraphics[width=\columnwidth]{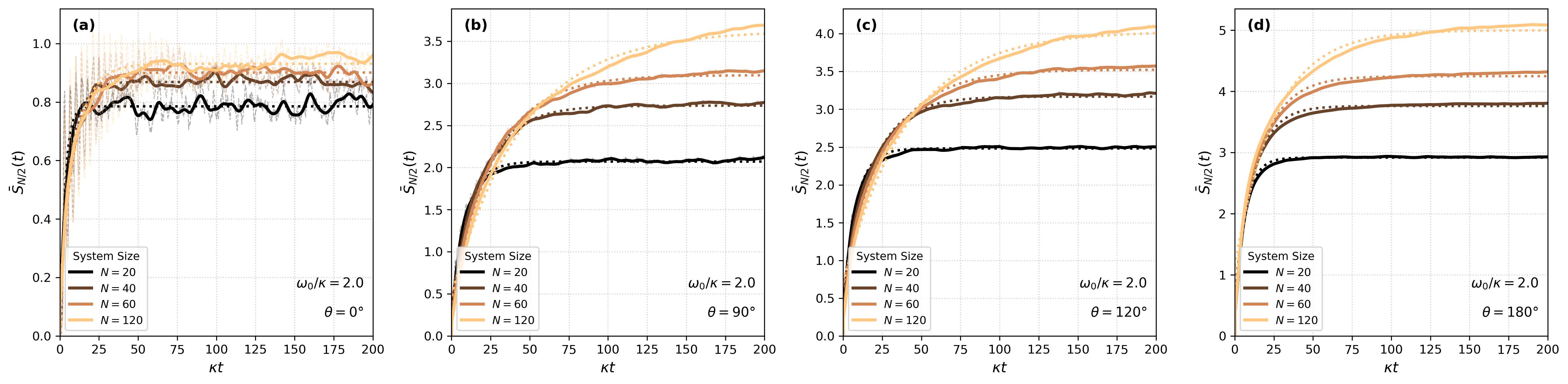}
\caption{\raggedright Entropy dynamics $\bar{S}_{N/2}(t)$ at
$\omega_0/\kappa=2.0$ for $\theta=0^\circ, 90^\circ, 120^\circ, 180^\circ$
(panels a--d) and $N=20, 40, 60, 120$. Transparent lines: raw
trajectory-averaged traces. Solid lines: Gaussian-filtered data, shown as
a guide to the eye. Dotted lines: exponential fit Eq.~\eqref{eq:exp_fit}
to the raw trace over the full simulation window. The entropy saturates
within the window for all system sizes, except $N=120$ for
$\theta\geq90^\circ$.}
\label{fig:entropy_dynamics}
\end{figure}

\twocolumngrid

The saturation time $t^*$ is extracted by fitting the trajectory-averaged
entanglement entropy $\bar{S}_{N/2}(t)$ to the functional form
\begin{equation}
\bar{S}_{N/2}(t) = a + b\,e^{-\kappa t/t^*},
\label{eq:exp_fit}
\end{equation}
where $a$, $b$, and $t^*$ are free parameters, the same functional form used in Ref.~\cite{Passarelli2024_ManybodyDynamics}. The fit is a least-squares fit to the raw trajectory-averaged trace, downsampled to about $2000$ points, over the full simulation window $\kappa t\in[0,200]$. The asymptotic value $a$ is initialized from the mean of the last 50 points of the trace, $b$ from $\bar{S}_{N/2}(0)-a$, and $t^*$ from $t_{\rm final}/5 = 40\,\kappa^{-1}$.

Figure~\ref{fig:entropy_dynamics} shows the entropy dynamics $\bar{S}_{N/2}(t)$ for all four angles at $\omega_0/\kappa=2.0$ and $N=20, 40, 60, 120$. Transparent lines are the raw traces, solid lines are Gaussian-filtered data shown as a guide to the eye, and dotted lines show the fit~\eqref{eq:exp_fit}, which gives the values of $t^*$ in Fig.~\ref{fig.ent.dynamics.saturation}(c). For $\theta=0^\circ$, and for $\theta\geq90^\circ$ with $N\leq60$, the entropy saturates within the window and the fit reliably extracts $t^*$. For $\theta\geq90^\circ$ and $N=120$, the entropy continues to grow throughout the window, so the fit follows the slow growth rather than a plateau. The returned $t^*$ (for example $\kappa t^*\approx 26$ at $\theta=\pi$) marks where the growth begins to slow and should be read as a lower bound on the true saturation time.

\end{document}